\documentclass[journal=jacsat,manuscript=article]{achemso}
\usepackage[version=3]{mhchem} 
\usepackage{xspace}
\usepackage[pdftex]{hyperref}
\usepackage{graphicx}
\usepackage[dvipsnames]{xcolor}

\newcommand{\celsius}{$^{\circ}$C\xspace}
\newcommand{\sio}{SiO$_\mathrm{x}$\xspace}
\newcommand{\sn}{SiN$_\mathrm{x}$\xspace}
\newcommand{\gao}{GaO$_\mathrm{x}$\xspace}

\newcommand{\LE}{lateral etching\xspace}
\newcommand{\Oxy}{$\mathrm{O}_2$\xspace}
\newcommand{\DSR}{diameter--spacing ratio\xspace}
\author{Jingxuan Kang}
\affiliation[PDI]
{Paul-Drude-Institut für Festkörperelektronik, Leibniz-Institut im Forschungsverbund Berlin e.V., Hausvogteiplatz 5--7, 10117 Berlin, Germany.}
\author{Rose-Mary Jose}
\affiliation[KIT]
{Paul-Drude-Institut für Festkörperelektronik, Leibniz-Institut im Forschungsverbund Berlin e.V., Hausvogteiplatz 5--7, 10117 Berlin, Germany.}
\altaffiliation
{Present address: Institute of Nanotechnology, Karlsruhe Institute of Technology, Hermann-von-Helmholtz-Platz 1, 76344 Eggenstein-Leopoldshafen, Germany}
\author{Oliver Brandt}
\affiliation[PDI]
{Paul-Drude-Institut für Festkörperelektronik, Leibniz-Institut im Forschungsverbund Berlin e.V., Hausvogteiplatz 5--7, 10117 Berlin, Germany.}
\author{Lutz Geelhaar}
\email{geelhaar@pdi-berlin.de}
\affiliation[PDI]
{Paul-Drude-Institut für Festkörperelektronik, Leibniz-Institut im Forschungsverbund Berlin e.V., Hausvogteiplatz 5--7, 10117 Berlin, Germany.}

\title{Combining metal dewetting and lateral etching for the scalable top-down fabrication of GaN nanowire arrays with independently tunable diameter and spacing}

\abbreviations{IR,NMR,UV}

\begin{document}


\begin{abstract}

The top-down fabrication of nanowires based on patterning via metal dewetting is a cost-effective and scalable approach that is particularly suited for applications requiring large arrays of nanowires. Advantageously, the nanowire diameter can be tailored by the initial metal film thickness. However, we show here that metal dewetting inherently leads to a coupling between the nanowire diameter and spacing. To overcome this limitation, we introduce two strategies that are exemplified for GaN nanowires: (i) modification of the surface and interface energies within the dewetting system, and (ii) thinning of the nanowires by \LE. In the first strategy, GaN(0001), \sio, and \sn substrate surfaces are combined with Au, Pt, and Pt–Au alloy dewetting metals to tune the dewetting behavior. The differences in interface energies affect the relation between nanowire diameter and spacing, albeit within a limited range. The second strategy adds a \LE step to the conventional top-down nanowire fabrication process. This step at the same time reduces the nanowire diameter and increases the spacing, thus enabling combinations beyond the constraints of metal dewetting alone. When in addition different initial nanowire diameters are employed, it is possible to independently control diameter and spacing over a substantially extended range. Therefore, the inherent limitation of conventional dewetting-based patterning approaches for the top-down fabrication of nanowires is overcome.

\end{abstract}

\section{Introduction}
Semiconductor nanowires have attracted considerable research interest owing to their conceptual advantages over planar thin films with respect to structural and optical properties.\cite{Garnett_2019} In particular, nanowires have been explored for a variety of applications, including light-emitting diodes (LEDs), photodetectors, and photoelectrochemical devices.\cite{Sang_2013,zhao2015,Li_2023,andrei2023,vignesh2024} The synthesis of nanowires can generally be categorized into two main approaches: bottom-up and top-down.\cite{hobbs2012a,mcintyre2020a} Compared to bottom-up techniques, top-down approaches offer better control over nanowire morphology and composition, and are thus widely utilized in the aforementioned applications.\cite{Bai_2012,Wang_2014,Jiao_2016,Behzadirad_2018,Gibson_2019,Narangari_2017} Among the various patterning techniques employed in the top-down approach, metal dewetting---a process in which a continuous metal film spontaneously breaks up into nanoscale islands upon thermal treatment---stands out due to its simplicity and scalability, offering sub-50\,nm resolution over large areas.\cite{liu2013a,yiyu2016a,kang2024b} The dimensions of the resulting patterns can be tuned by varying the initial metal film thickness and the thermal budget applied during dewetting.\cite{Kargupta_2001,Trice_2007,leroy2016,thompson2012a,yiyu2016a,kang2024b,Ruffino_2019}

Here, we identify a fundamental limitation of metal dewetting for top-down nanowire fabrication that has received little attention so far. Specifically, the mean diameter of the metal nanoislands and their average spacing are inherently coupled due to the nature of metal dewetting. Although both parameters have been observed to scale with the initial metal film thickness,\cite{Kargupta_2001,Trice_2007, leroy2016,thompson2012a,yiyu2016a,kang2024b,Ruffino_2019} to the best of our knowledge the implications of this correlation have not been discussed in the context of nanowire fabrication. This interdependence presents a challenge for nanowire applications requiring independent control of nanowire diameter and spacing, as enabled by, e.\,g., electron beam lithography. For instance, the optical coupling efficiency in photonic devices is strongly influenced by both nanowire diameter and pitch.\cite{anttu2013b,wallentin2013,Anttu_2020} Moreover, insufficient spacing may lead to shadowing effects during radial shell growth on nanowires, thereby affecting the shell structure and composition.\cite{Czaban_2009,Al_Humaidi_2023}

In the present work, we demonstrate two distinct approaches to overcome this constraint, using GaN nanowires as a model system. GaN nanowires are of particular interest due to their advantages for optoelectronic and photoelectrochemical applications.\cite{Bai_2012,Wang_2014,zhao2015,Narangari_2017,Hou_2019,Zhao_2024,vignesh2024} In the first approach, the dewetting material system is modified in order to alter the surface and interface energies governing the dewetting dynamics. The second approach is based on \LE,\cite{oliva2023a} which increases the spacing between nanowires by reducing their diameter. The latter method enables the independent tuning of diameter and spacing within a substantial range. Therefore, the versatility of metal dewetting for nanowire-based device fabrication is significantly enhanced.

\section{Results and discussion}

\begin{figure}[t]
	\centering
	\includegraphics[width=0.75\columnwidth]{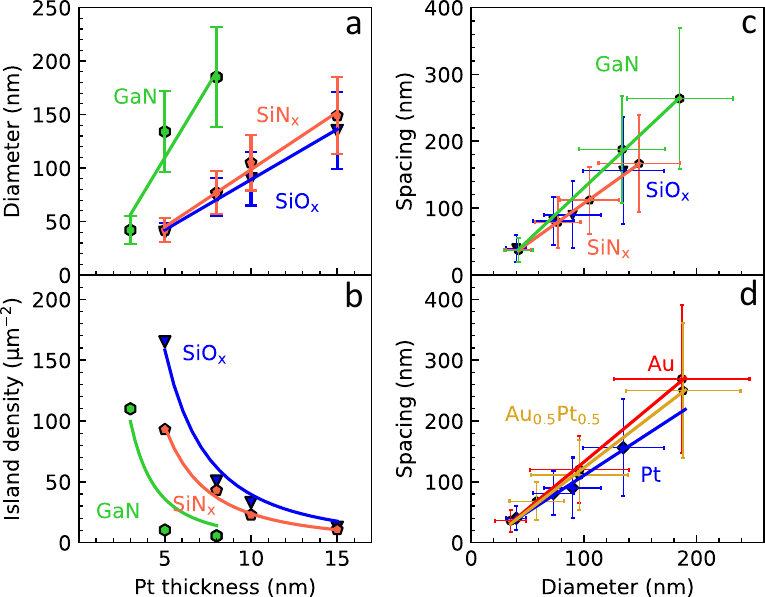}
	\caption{(a) and (b) show the dependence of nanoisland diameter and density on Pt film thickness for different dewetting substrates, as indicated by the text labels. (c) and (d) show correlations of the dewetting-formed nanoislands' mean diameter and the average spacing between nanoislands for different substrates and dewetting metals, respectively. Data points correspond to experiment results, and the solid lines are the result of fits as described in the main text. }
	\label{fig:dewetting nature}
\end{figure}

We investigate the annealing-induced dewetting of thin Pt films with thicknesses between 5 and 15\,nm deposited on GaN(0001), \sio, and \sn surfaces. Figures~\ref{fig:dewetting nature}~(a) and (b) display the mean diameter and density of the resulting nanoislands. In all cases, the mean diameter increases approximately linearly with increasing Pt film thickness, while the density decreases inversely quadratically, as indicated by the solid lines. We have reported these trends in our previous work,\cite{kang2024b} and they are in agreement with earlier studies.\cite{Capellini_2009,leroy2016} A more detailed quantitative discussion is provided below. We note that we cite here review publications revealing that such dependencies of nanoisland diameter and density on initial film thickness have been widely reported for metal dewetting in general, not only in relation to top-down nanowire fabrication.  From the same set of samples, we also extracted the average spacing between adjacent nanoislands, as shown in Fig.~\ref{fig:dewetting nature}~(c). The solid lines show that the mean diameter and spacing exhibit a clear linear correlation for all three substrates, indicating a coupling between these two parameters. Hence, if such ensembles of nanoislands formed by dewetting are used for the top-down fabrication of nanowires, their diameter and spacing cannot be chosen independently. This restriction constitutes the central challenge addressed in the present work.


Since both the nanoisland diameter and density depend on the initial film thickness, it is not surprising that the former two parameters are coupled. Now, we consider this coupling in more detail. Given that in the current study the annealing was carried out in a tube furnace for durations exceeding one hour, one can assume that the resulting nanoisland ensemble is close to the equilibrium state. Thus, we employ a simplified thermodynamic model for solid film dewetting.\cite{leroy2016} 
This model neglects the mechanisms of nanoisland formation and kinetic effects. Under near-equilibrium conditions, the minimum island diameter $d_{\min}$ and the corresponding island density $\rho$ for a given dewetting film thickness $h_A$ are expressed as:

\begin{equation}
d_{\min} = 3K h_A
\label{eq:d_min}
\end{equation}

\begin{equation}
\rho = \frac{4}{9 K^3 \pi g(\theta)} h_A^{-2}
\label{eq:rho}
\end{equation}
Here, A denotes the metal film and B the substrate. The factor $K$ is defined as 
\[
K = \frac{2 f(\theta) \gamma_{A} + \gamma_{AB} - \gamma_{B}}{g(\theta) (\gamma_{A} - \gamma_{B} + \gamma_{AB})},
\]
where the functions $f(\theta)$ and $g(\theta)$ are given by
\[
f(\theta) = \frac{1 + \cos \theta}{\sin^2 \theta}, \quad
g(\theta) = \frac{1 + \cos \theta}{2} \sin^3 \theta \left[1 + \cos \theta + \sin^2 \theta \right].
\]
$\theta$ is the contact angle of the nanoislands on the substrate. The parameters $\gamma_A$, $\gamma_B$, and $\gamma_{AB}$ are the surface energies of materials A and B and the corresponding interface energy, respectively. Since $\theta$ also depends on these energies, the size and density of the resulting nano-islands are governed by the energetics of the interface. Figs.~\ref{fig:dewetting nature}~(a) and (b) show that these equations well describe our experimental data, as previously reported.\cite{kang2024b}

From the above equations, one can derive the following expression for the ratio between the spacing $l$ and the minimum island diameter $d_{\min}$:

\begin{equation}
\frac{l}{d_{\min}} = \sqrt{ \frac{\pi}{4} \left[ \frac{2 f(\theta) \gamma_A - \gamma_B + \gamma_{AB}}{\gamma_A - \gamma_B + \gamma_{AB}} \right] } - 1
\label{formula_spacing and diameter}
\end{equation}

The derivation of this expression for the \DSR  is provided in the Supplementary Information. Indeed, for a given dewetting system the diameter--spacing ratio is fixed as suggested by the solid lines in Figs.~\ref{fig:dewetting nature}~(c). Furthermore, the ratio is solely determined  by the material-specific energies $\gamma_A$, $\gamma_B$, and $\gamma_{AB}$.

An obvious option to modify  the \DSR is the adjustment of either material A or B and, thus, $\gamma_A$ or $\gamma_B$ and $\gamma_{AB}$. For the data in Fig.~\ref{fig:dewetting nature}~(c), material A was kept constant (Pt) and material B was varied [GaN(0001), \sio, and \sn]. The different slope observed for GaN(0001) compared to \sio and \sn reveals a modification of the \DSR arising from differences\cite{Zhang_2016, Shklyaev_2020} in surface and interfacial energies. At the same time, the slopes for the two substrate materials \sio and \sn are nearly identical. This similarity likely stems from the fact that both substrates are amorphous, whereas GaN(0001) is crystalline. In any case, this comparison reveals that changes in substrate composition may not suffice to modify the \DSR. Furthermore, practical constraints may also limit the applicability of adjusting the substrate material. For example, the thermal stability during annealing, the etching rate during subsequent top-down nanowire fabrication, and the deposition cost are important criteria for the material selection. 

Besides modifying the substrate material B, the metal film material A can also be changed to alter the dewetting behavior. This approach offers more flexibility as metals can typically more easily be alloyed than non-metallic substrate materials, thus providing a continuous variation in surface and interface energies. Furthermore, nanowire top-down processing imposes fewer constraints on the choice of the dewetting metal, since the metal does not directly contact the target nanowire material and can be removed by selective etching of the buffer layer.

We demonstrate this concept using Pt, Au, and a Pt–Au alloy with a 1:1 composition ratio. Fig.~\ref{fig:dewetting nature}~(d) displays the corresponding nanoisland spacing and diameter, along with linear fits. Notably, the slopes differ markedly between pure Pt and Au, while the alloy exhibits intermediate behavior. Hence, the \DSR depends on alloy composition. By adjusting the ratio of Pt to Au in the alloy, the slope---and hence the \DSR---can be tuned, highlighting the potential of alloying for controlling the dewetting behavior.

\begin{figure*}
    \includegraphics[width=\linewidth]{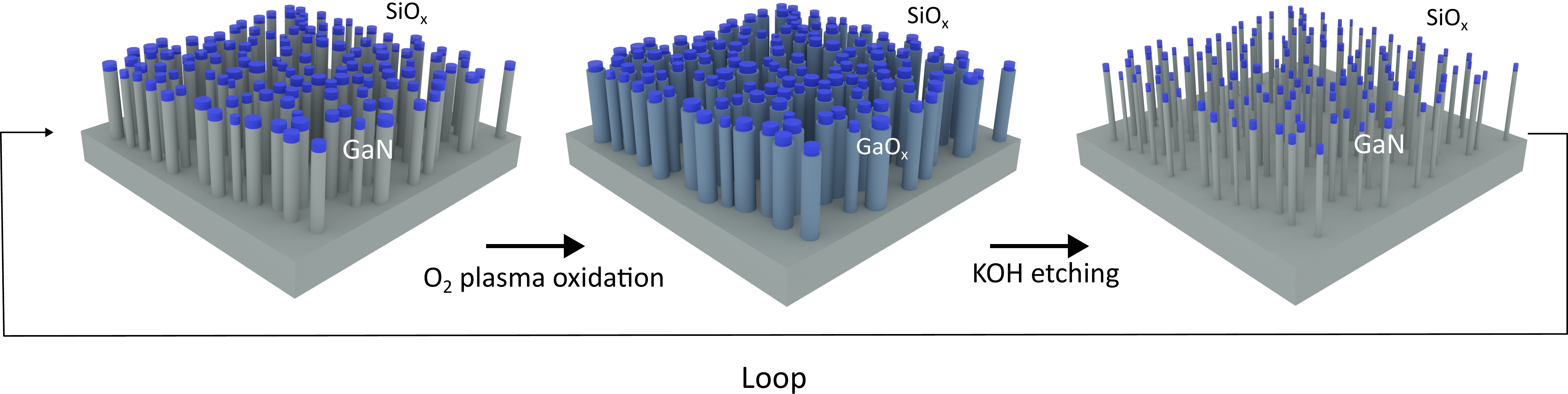}
    \caption{Step flow diagram and schematics for the lateral etching of top-down GaN nanowires.}
    \label{step flow}
\end{figure*}

Figures~\ref{fig:dewetting nature}~(c) and (d) show that varying the dewetting metal and substrate can yield different nanoisland diameter--spacing ratios. However, the overall tunability remains limited. To overcome the constraints inherent to the process of metal dewetting and widen the choice of nanowire diameter--spacing combinations, we consider the entire process of top-down nanowire fabrication and introduce the \LE method. The underlying principle of this approach is the reduction of the nanowire diameter to increase the spacing. Fig.~\ref{step flow} presents a schematic illustration of the \LE process. Initially, a GaN nanowire ensemble capped with a \sio buffer layer is fabricated using the nanoislands resulting from metal dewetting as an etching mask. The GaN nanowire arrays are subsequently subjected to an \Oxy plasma treatment to oxidize their sidewalls. The resulting \gao layer is then selectively removed by immersing the sample in a KOH solution. This sequence increases the spacing between nanowires by reducing their diameter and can be repeated for the same array. By selecting appropriate initial metal film thicknesses and \LE parameters, nanowire ensembles with tailored diameters and spacings can be fabricated.

Accurate control of the \gao thickness is critical for precisely tuning the nanowire dimensions in the \LE process. We note that our group previously developed this process to obtain ultrasmall diameters for top-down GaN nanowires fabricated by means of electron-beam lithography.\cite{oliva2023a} Besides the difference in mask fabrication, that processing scheme involved other materials. To examine the  \LE of GaN nanowires fabricated by means of metal dewetting, we performed a series of oxidation experiments with RF powers between 30 and 100\,W and oxidation times between 2 and 15\,min. The \gao thickness formed on the nanowire surface was estimated from secondary electron (SE) micrographs by measuring the mean nanowire diameter before and after \LE. The \gao thickness was deduced from the average diameter difference (see Supporting Information for details). It should be noted that after oxidation, the nanowire diameter increases slightly due to the various factors, most notably the larger specific volume of the \gao.\cite{oliva2023a} Here, we ignore this phenomenon as the corresponding difference in diameter is too small to be precisely determined by our measurements. In addition, the heated KOH solution may also etch GaN and thus reduce the nanowire diameter. However, the lateral etching rate on the $M$-plane is slow, and the etching duration was limited to 3\,min. Therefore, the effect of KOH on GaN during this process is negligible.

In general, the oxide thickness resulting from the plasma oxidation of semiconductor nanowires is influenced by several factors, including temperature, oxidation time, the partial pressure of oxygen, applied power, and nanowire diameter.\cite{Nagai_2003,buttner2006,Tinoco_2006,Krzeminski_2012,liu2016,itoh2021,Oon_2013,Sokolovskij_2016,Bui_2019,oliva2023a} Only the last parameter is specific to the nanowire geometry. In particular, the nanowire sidewall curvature ($1/r$) imposes a limitation on the oxidation rate, commonly referred to as the self-limiting effect.\cite{buttner2006,liu2016} Two physical phenomena have been proposed to explain this effect. As oxidation takes place, radial stress builds up at the interface between semiconductor and oxide. First, this stress can limit the oxidation rate. Second, the the stress may increase the activation energy for oxidant diffusion, thereby limiting the supply of oxidant species to the interface.\cite{Krzeminski_2012,liu2016,itoh2021} 

Quantitatively, we assume on the basis of more detailed studies\cite{Tinoco_2006,itoh2020}  that the oxide thickness $x_{\mathrm{o}}$ resulting from plasma treatment follows a power-law dependence:
\begin{equation}
    x_{\mathrm{o}} = K \, \alpha(T, P_{\mathrm{power}}, p_{\mathrm{O_2}}) \, t^{\beta},
\end{equation}
where $t$ is the oxidation duration and the function $\alpha(T, P_{\mathrm{power}}, p_{\mathrm{O_2}})$ encompasses the effects of oxidation temperature, plasma power, and partial pressure of oxygen. $K$ and $\beta$ are constants. In this work, $\alpha(p_{\mathrm{O_2}})$ is also constant as the plasma was generated under fixed oxygen flow and chamber pressure. In addition, we  assume that the oxidation temperature remained constant across all experiments since it was not intentionally varied. 

Following the power-law model,\cite{Tinoco_2006,itoh2020} $\alpha(P_{\mathrm{power}})$ can be expressed as
\begin{equation}
    \alpha(P_{\mathrm{power}}) = C \, P_{\mathrm{power}}^{q},
\end{equation}
where $C$ and $q$ are constants. Substituting this expression into the general power-law form yields the relation
\begin{equation}
    x_{\mathrm{o}}(P_{\mathrm{power}}, t) = K \, P_{\mathrm{power}}^{q} \, t^{\beta}.
    \label{eq: CP4_LE: oxidation thickness}
\end{equation}

\begin{figure}
    \centering
    \includegraphics[width=0.5\linewidth]{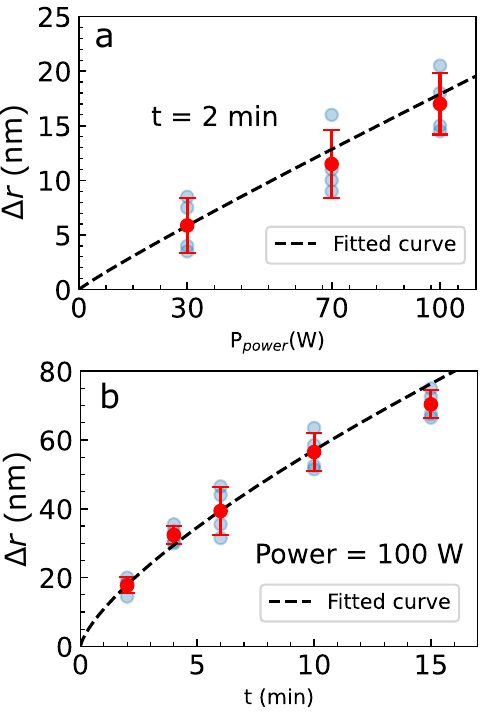}
    \caption{(a) Reduction in the mean radius of GaN nanowires as a function of RF power for a fixed exposure time of 2\,min. (b) Reduction in the mean radius of GaN nanowires via \LE as a function of oxidation time (2--15\,min) at a fixed RF power of 100\,W. The dashed lines represent fits to equation~\ref{eq: CP4_LE: oxidation thickness}. Light blue points denote the individual measured values from nanowire samples, while red points indicate the corresponding mean values with error bars.}
    \label{fig:CP4_LE: oxidation thickness}
\end{figure}

Figure~\ref{fig:CP4_LE: oxidation thickness} presents the experimental data for the mean reduction in radius $\Delta r$ together with fits based on equation~\ref{eq: CP4_LE: oxidation thickness}. The fitting results in $K = 0.15 \pm 0.04$, $q = 0.93 \pm 0.07$, and $\beta = 0.72 \pm 0.04$. The coefficient of determination is $R^2 = 0.99$, indicating a good agreement between the model and the experimental data. As seen in Fig.~\ref{fig:CP4_LE: oxidation thickness}~(a), for a fixed oxidation time, $\Delta r$ and thus the \gao thickness increase with RF power, which can be attributed to the higher density and/or higher kinetic energy of oxidizing radicals in the plasma. We note that the value of $q$ is close to one, consistent with the simpler linear dependence that we assumed in our previous publication.\cite{oliva2023a} In contrast, Fig.~\ref{fig:CP4_LE: oxidation thickness}~(b) shows that at constant RF power, $\Delta r$ and thus the \gao thickness increase with oxidation time with a dependence that is for short times clearly non-linear. Hence, the oxidation rate, which is the derivative of $\Delta r$ with respect to $t$, decreases with increasing time, consistent with the self-limiting effect discussed above.\cite{buttner2006,liu2016} The close agreement between fits and experimental data confirms that the oxidation model accurately describes the oxidation behavior of GaN nanowires. These results demonstrate that the diameter of GaN nanowires can be effectively tuned by the \LE process.

\begin{figure}
	\centering
	\includegraphics[width=\columnwidth]{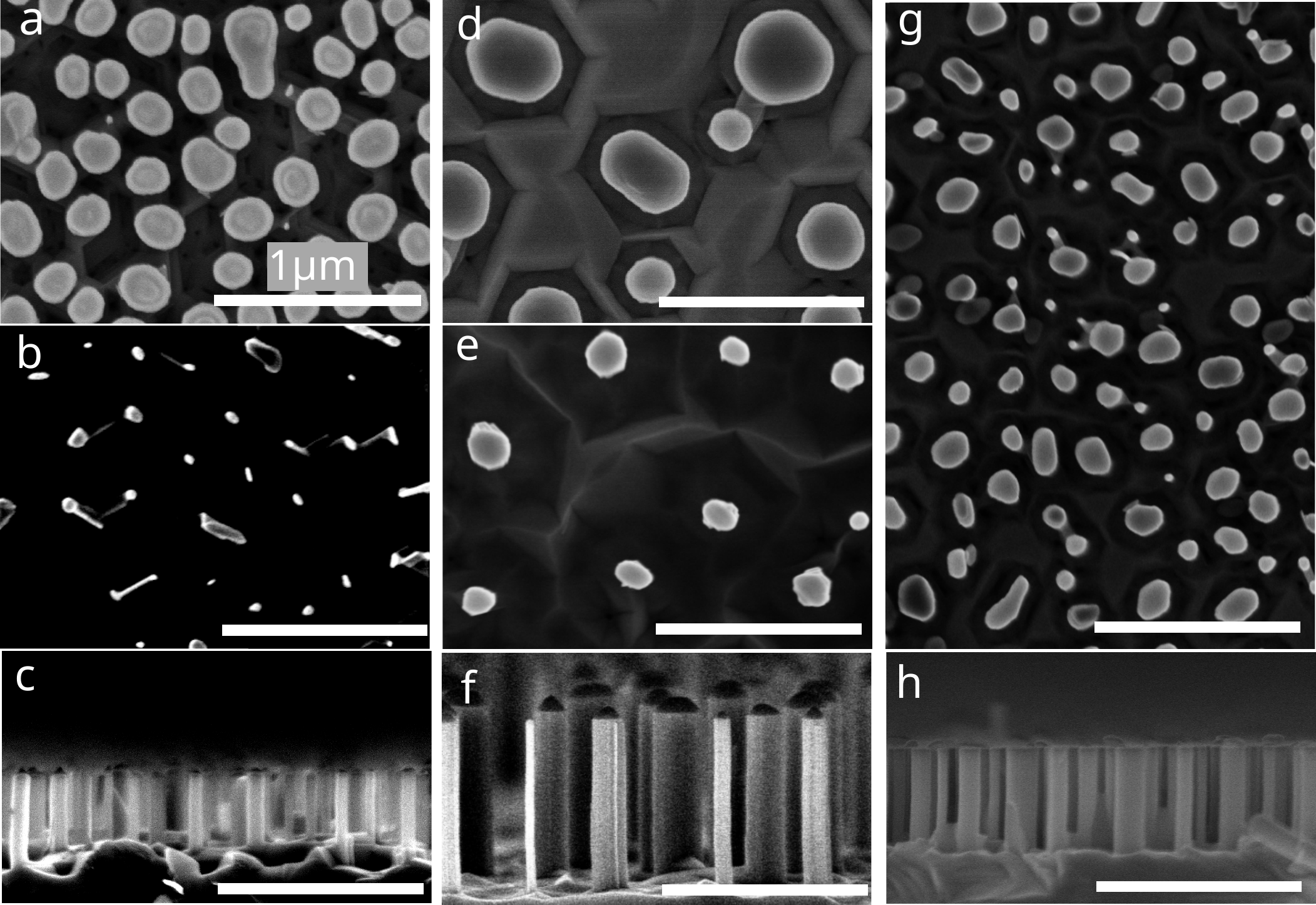}
	\caption{SE micrographs of top-down GaN nanowires prepared by metal dewetting and subsequent lateral etching. (a) Top-view of nanowires from a 15\,nm Pt film (mean diameter $\approx$190\,nm, spacing $\approx$170\,nm). (b) Top-view and (c) side-view of the same nanowire ensemble after \LE (mean diameter $\approx$70\,nm, spacing $\approx$380\,nm). (d) Top-view of nanowires from a 12\,nm Au film (mean diameter $\approx$310\,nm, spacing $\approx$430\,nm). (e) Top-view and (f) side-view of the same nanowire ensemble after \LE (mean diameter $\approx$170\,nm, spacing $\approx$750\,nm). (g) Top-view and (h) side-view of a reference nanowire ensemble from a 10\,nm Pt film without \LE that exhibits similar diameter but very different spacing compared to the sample shown in (e) and (f) (mean diameter $\approx$140\,nm, spacing $\approx$190\,nm).}   
	\label{LE samples}
\end{figure}

In order to demonstrate the effect of \LE on the \DSR, two GaN nanowire ensembles were fabricated and subsequently subjected to \LE. These samples were initially prepared by dewetting 15\,nm-thin Pt and 12\,nm-thin Au films on \sio buffer layers, respectively. Both samples then underwent \LE at an RF power of 100\,W for 15\,min. SE micrographs of the initial and final nanowire ensembles are depicted in Figs.~\ref{LE samples}(a)--(f). Before \LE, the mean diameters were $\approx$190\,nm and $\approx$310\,nm, with corresponding average spacings of $\approx$170\,nm and $\approx$430\,nm. Following \LE, the mean diameters were reduced to $\approx$70\,nm and $\approx$170\,nm, while the average spacings increased substantially to $\approx$380\,nm and $\approx$750\,nm, respectively. Notably, the reduction in diameter was smaller than the increase in spacing. This discrepancy is attributed to the removal of the thinnest nanowires during the KOH etching step, which reduces the nanowire density and thereby increases the average spacing. The removal of these thin nanowires is not due to their vanishing, but rather due to their mechanical instability; ultrathin nanowires tend to collapse by lateral bundling or breaking, as reported previously.\cite{oliva2023a}

Next, we illustrate how the combination of different initial dewetting parameters and \LE can be employed to obtain nanowire arrays that exhibit similar diameters but clearly differ in average spacing.  Figs.~\ref{LE samples}~(g) and (h) present top-view and side-view SE micrographs of a reference nanowire ensemble fabricated from a 10\,nm Pt dewetting pattern, exhibiting a mean diameter of $\approx$140\,nm and an average spacing of $\approx$190\,nm. This sample is to be compared with the one shown in Figs.~\ref{LE samples}~(e) and (f), i.~e. the laterally etched nanowire ensemble derived from the 12\,nm Au dewetting sample. Those nanowires exhibit a mean diameter similar to that of the reference sample, but with a markedly larger average spacing. These results demonstrate that \LE provides an effective means to decouple the diameter and spacing in top-down GaN nanowire arrays fabricated on the basis of metal dewetting.

\begin{figure}[t]
	\centering
	\includegraphics[width=0.5\columnwidth]{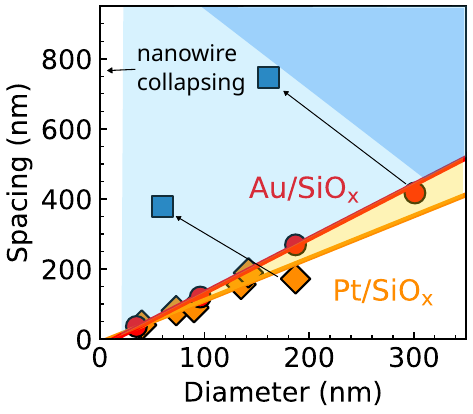}
	\caption{Combinations of GaN nanowire diameter and spacing obtained by top-down fabrication based on metal dewetting on a \sio buffer layer. The symbols indicate experimental values for the dewetting of  Pt (diamonds) and Au (spheres). The solid lines correspond to linear fits. The square-shaped symbols depict experimental values resulting from \LE for 15\,min with an RF power of 100\,W, as marked by the arrows. The yellow-shaded area is accessible by changing the Au-Pt alloy composition. The blue-shaded region is unlocked by \LE, with the deep blue-shaded region requiring larger nanowire diameter than in the samples fabricated for the present study.}
	\label{etching summary}
\end{figure}

Fig.~\ref{etching summary} summarizes the combinations of spacing and diameter obtained by different approaches. Varying the dewetting metal (Au, Pt, and their alloys) yields the yellow-shaded range. For simplicity, we focus on dewetting on \sio buffer layers and have left out the change in substrate material since it does not unlock other diameter--spacing combinations. The accessible parameter space is substantially extended to the blue-shaded region above the 'Pt/\sio' line by \LE. This region does not include very small diameters since ultrathin nanowires resulting from \LE may bundle or break off.\cite{oliva2023a} Furthermore, since \LE can only reduce nanowire diameters, the region below the 'Pt/\sio' line remains inaccessible. This limitation could be overcome by finding dewetting systems with smaller slopes or by epitaxial shell growth that enlarges the diameter.\cite{Dagher_2019,van_Treeck_2023}

\section{Conclusions}

We have identified an inherent limitation of nanoscale patterning based on metal dewetting--the diameter and spacing of the resulting nanoislands cannot be controlled independently. The coupling of these two parameters restricts the patterning parameter range that can be accessed by this simple technique and thus its usability for top-down nanowire fabrication \cite{liu2013a,yiyu2016a,kang2024b,Narangari_2017} and other nanoscale systems such as bottom-up nanowire growth.\cite{Yang_2002,Hannon_2006,Wang_2006} To overcome this limitation in the concrete case of GaN nanowires, we have introduced two strategies: modification of the dewetting material system and the \LE of nanowires. The first strategy widens the parameter range but only by a small extent. In contrast, \LE enables the independent tuning of GaN nanowire diameter and spacing in a substantial range. Hence, the inherent limitation of metal dewetting is mitigated. Furthermore, the combination of metal dewetting patterning with \LE can readily be transferred to other semiconductor nanowires, such as Si, since the necessary etching protocols are known or can be developed for a wide range of materials. Therefore, the strategies presented here make metal dewetting significantly more useful for the wafer-scale patterning of nanowire ensembles at comparatively low fabrication cost, which is attractive for various applications.

\section{Experimental}

Decoupling the diameter and spacing of nanoislands by modifying the dewetting materials involved adjusting both substrate and dewetting metal. Specifically, the dewetting substrates included a commercial 5-$\mu$m-thick single-crystalline GaN(0001) template grown by metal organic vapor phase epitaxy on a single-crystalline sapphire substrate, as well as amorphous \sio (thickness 220 nm) and \sn (20 nm) layers sputtered onto the GaN(0001) template. The dewetting metals Pt, Au, and a 1:1 Pt–Au alloy were deposited with thicknesses between 5 and 15 nm using electron beam evaporation. The metal dewetting process was carried out in a tube furnace under an Ar atmosphere, with the temperature being ramped up at a rate of 25\,\celsius/min, followed after the chosen annealing time by cooling at the rate resulting from heat dissipation to the lab environment. Annealing conditions were set to 900\,\celsius for 90\,min for Pt, Au, and Pt-Au alloy dewetting on \sio and \sn buffers, and 800\,\celsius for 60\,min for Pt dewetting on GaN(0001) templates. Both the annealing temperature and duration were reduced for the GaN(0001) templates to avoid decomposition. The dewetting experiments follow procedures established in our previous work.\cite{kang2024b} The statistical analysis of the resulting nanoisland dimensions was conducted by the image processing of SE micrographs involving both ImageJ and a Python code developed in-house. This image analysis is also detailed in our previous work.\cite{kang2024b} \par

For decoupling nanowire diameter and spacing via the \LE method, the principal mechanism relies on reducing the lateral size of nanowires through an oxidation–oxide removal sequence, a process previously applied by our group to fabricate thin nanowires.\cite{oliva2023a} Top-down GaN nanowire ensembles were fabricated using a combination of metal dewetting patterning, inductively coupled plasma dry etching, and KOH solution etching, as described in our earlier work.\cite{kang2024b} The fabricated nanowire ensembles subsequently underwent \LE, beginning with oxidation in an \Oxy plasma environment generated in a reactive ion etching (RIE) chamber. The oxidation process was carried out using an \Oxy flow of 10\,sccm at a chamber pressure of 7\,Pa, with durations between 2 and 15\,min and RF powers between 30 and 100\,W. During plasma oxidation, a \gao layer formed on the GaN nanowire side surfaces, with its thickness determined by the plasma parameters. Following oxidation, the samples were immersed in a 30\,wt\% KOH solution heated to 65\,\celsius for 3\,min to remove the \gao layer.\cite{li2011a,oliva2023a} The diameter, number density, and spacing of the GaN nanowires before and after \LE were evaluated from SE micrographs, as described in the Supplementary Information.

\begin{acknowledgement}

The authors express their gratitude to Claudia Herrmann for her technical assistance, to Sander Rauwerdink, Walid Anders, and Nicole Volkmer for assistance with nanowire sample processing, to Anne-Kathrin Bluhm for obtaining SE micrographs, and to Wenshan Chen for a critical reading of the manuscript. This research was funded by the Deutsche Forschungsgemeinschaft under grant Ge2224/6-1.

\end{acknowledgement}

\bibliography{Ref}

\end{document}


\maketitle

\section*{Supplementary Information}

\subsection*{1. Statistical analysis of nanoisland dimensions}

The statistical analysis of nanoisland dimensions was performed using top-view secondary electron (SE) micrographs. First, the software ImageJ was used to identify individual nanoislands by binarizing the brightness contrast relative to the substrate background, thereby extracting the area, position, and circularity of each nanoisland. These data were then analyzed using Python code developed in-house to compute the equivalent disc diameter and the nearest-neighbor distances between nanoislands. From these results, the mean diameter, average spacing, and their standard deviations were determined. A detailed description of this procedure is provided in the Supporting Information of our previous work.\cite{kang2024b}  

\subsection*{2. Diameter--spacing ratio and thermodynamic model}

\vspace{0.5cm} 

\noindent
\begin{minipage}{0.5\textwidth}
\raggedleft
  \includegraphics[width=0.7\linewidth]{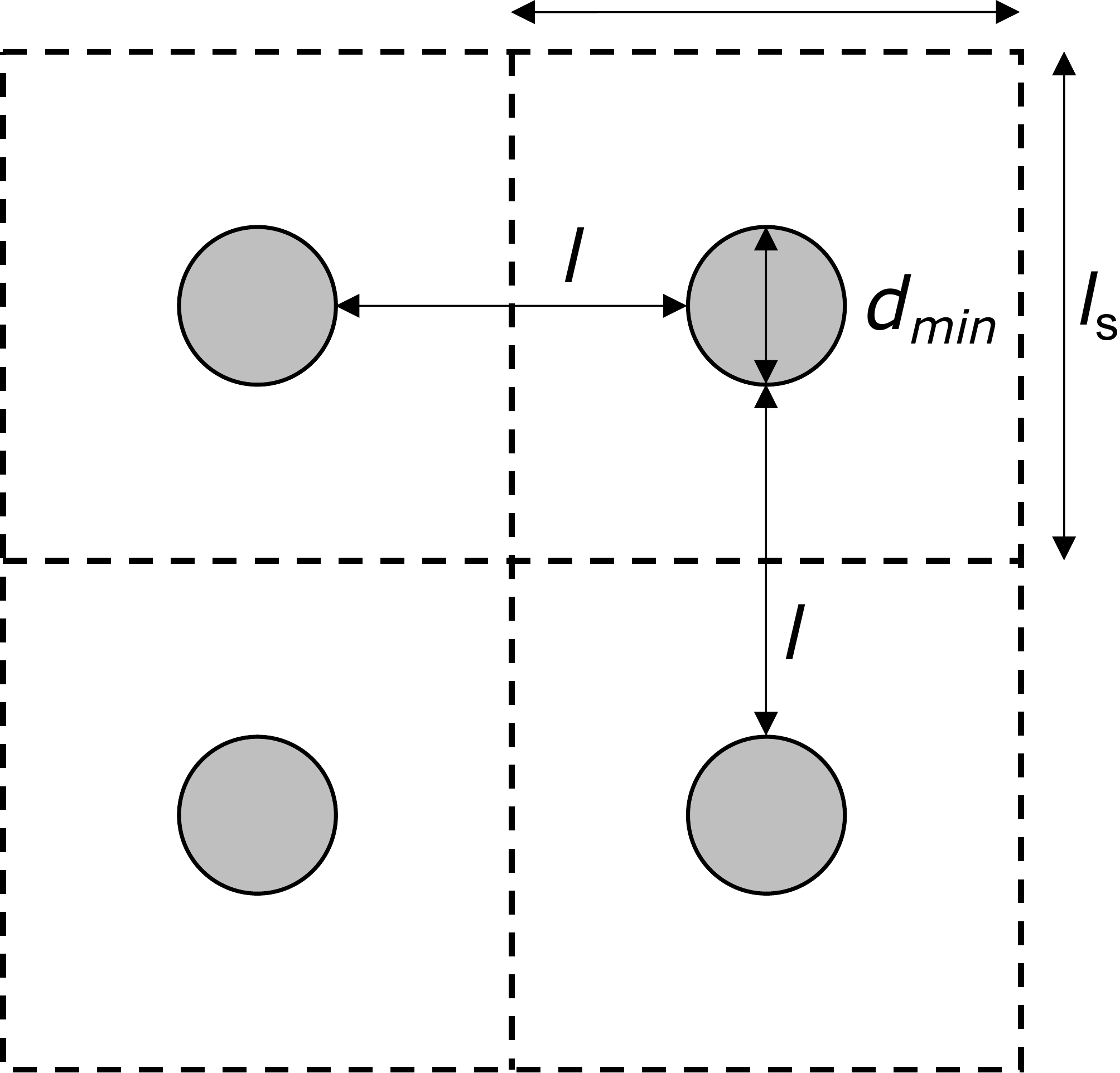}
\end{minipage}%
\hfill
\begin{minipage}{0.45\textwidth}
\raggedright
  \captionof{figure}{Schematic arrangement of uniformly spaced metal islands of equal diameter $d_{\min}$. The minimum separation distance between two adjacent islands is denoted as $l$. Each island can conceptually be associated with a square unit cell of side length $l_s$.}
  \label{SI: diameter spacing model}
\end{minipage}

\vspace{0.5cm} 

As discussed in the main paper, a thermodynamic model of metal dewetting describes under the assumption of thermal equilibrium the minimum diameter $d_{\min}$ and density $\rho$ of the resulting nanoislands.\cite{leroy2016}. On this basis, the ratio of spacing and diameter can be calculated from a simplified geometry, as illustrated in Fig.~\ref{SI: diameter spacing model}. The model assumes that the islands are spherical caps of equal size and uniformly distributed on the surface. Each island occupies a square unit cell of area  
\begin{equation}
    l_s^2 = \frac{1}{\rho},
\end{equation}
leading to
\begin{equation}
    l_s = \sqrt{\frac{1}{\rho}} = \sqrt{\frac{9 K^3 \pi g(\theta) h_A^2}{4}} = \frac{3}{2}h_A\sqrt{K^3 \pi g(\theta)}.
\end{equation}
The minimum distance $l$ between two neighboring islands is then
\begin{equation}
    l = l_s - d_{\min}.
\end{equation}
Thus,
\begin{equation}
    \frac{l}{d_{\min}} = \frac{l_s}{d_{\min}} - 1.
\end{equation}
Substituting $d_{\min} = 3K h_A$ and $K=\frac{2 f(\theta) \gamma_{A}+\gamma_{AB}-\gamma_{B}}{g(\theta)\left(\gamma_{A}-\gamma_{B}+\gamma_{AB}\right)}$ yields
\begin{equation}
    \frac{l}{d_{\min}} = \sqrt{\frac{\pi}{4}\left[\frac{2 f(\theta) \gamma_A-\gamma_B+\gamma_{AB}}{\gamma_A-\gamma_B+\gamma_{AB}}\right]} - 1.
\label{eq: formula_spacing and diameter}
\end{equation}

Equation~\ref{eq: formula_spacing and diameter} reveals that, for a given dewetting system, the diameter-to-spacing ratio is constant and is determined by $\gamma_A$, $\gamma_B$, and $\gamma_{AB}$, as $f(\theta)$ depends on these energies. In practice, nanoisland sizes and spacings are not perfectly uniform due to the complexity of the dewetting process. Therefore, the experimentally determined \emph{mean diameter} and \emph{average spacing} are used to represent $d_{\min}$ and $l$, respectively.

\subsection*{3. Determination of the oxidation thickness}

\begin{figure}[h]
\centering
\includegraphics[width=\columnwidth]{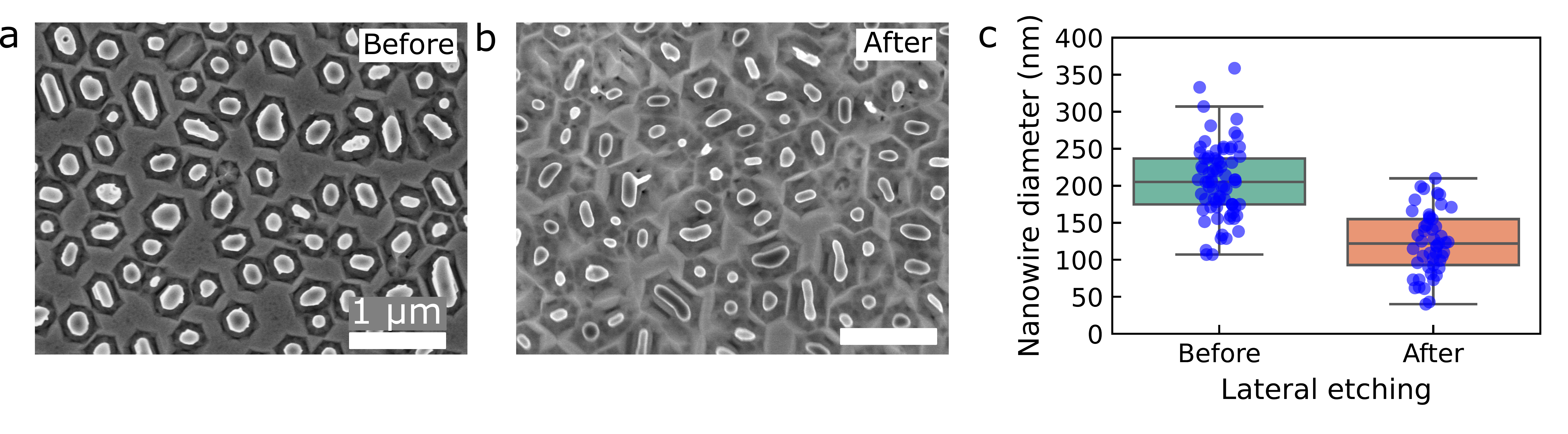}
\caption{(a), (b) Top-view SE micrographs of a GaN nanowire ensemble before and after \LE using for the oxidation 100\,W RF power for 4\,min. (c) Nanowire diameter distributions obtained from the image analysis of (a) and (b), the blue scatter data points are collected experiment data and the central line inside the box represents the median. The box edges mark the interquartile range (IQR) that contains the middle 50\% of the data, and the vertical bars extend up to 1.5 $\times$ IQR from the quartiles.}
\label{SI-figure: LE-SEM}
\end{figure}

The \gao thickness $d_\mathrm{GaO_x}$ of GaN nanowires oxidized in an \Oxy plasma was determined from top-view SE micrographs. The mean nanowire diameter before \LE, measured as described in Section~1, is denoted as $d_{b}$. After \LE, the mean diameter is measured again and denoted as $d_{a}$. Assuming complete removal of the \gao layer during KOH etching, the oxide thickness is
\begin{equation}
d_\mathrm{GaO_x} = d_{b} - d_{a}.
\label{eq: oxidation thickness}
\end{equation}
As discussed in the main paper, this procedure neglects that the nanowire diameter slightly increases after oxidation since the oxide thickness is larger than the thickness of the original GaN layer that is oxidized.

Figures~\ref{SI-figure: LE-SEM}~(a) and (b) show SE micrographs of GaN nanowires fabricated from the dewetting of an 8-nm-thin Pt layer on GaN(0001)  before and after the \LE procedure. The significant reduction in nanowire diameter is quantitatively shown in Fig.~\ref{SI-figure: LE-SEM}(c) based on the image analysis described above.

\bibliography{Ref}